\documentclass[iop]{emulateapj}
\usepackage{epsfig}
\usepackage{graphicx}
\usepackage{dcolumn}
\usepackage{bm}
\usepackage[colorlinks=true,linkcolor=blue,citecolor=blue]{hyperref}
\usepackage{natbib}
\usepackage{comment}
\usepackage{xcolor}
\usepackage{enumerate}
\usepackage{bigints}
\usepackage{multirow}
\usepackage{booktabs} 
\usepackage{mathtools}
\usepackage{makecell,tabularx}
\usepackage[normalem]{ulem}
\setcellgapes{3pt}
\graphicspath{{./}{figures/}}
\def\msun{\,\rm M_\odot}

\begin{document} 

\title{Filling the mass gap: How kilonova observations can unveil the nature of the compact object merging with the neutron star}

\author{C. Barbieri$^{1,3*}$,
          O.~S.~Salafia$^{2,3}$, M.~Colpi$^{1,3}$, G.~Ghirlanda$^{2,3}$, A.~Perego$^{3,4}$  and
          A.~Colombo$^1$}

\altaffiltext{1}{Università degli Studi di Milano-Bicocca, Dipartimento di Fisica ``G. Occhialini'', Piazza della Scienza 3, I-20126 Milano, Italy}
\altaffiltext{2}{INAF -- Osservatorio Astronomico di Brera, via E. Bianchi 46, I-23807 Merate, Italy}
\altaffiltext{3}{INFN -- Sezione di Milano-Bicocca, Piazza della Scienza 3, I-20126 Milano, Italy}
\altaffiltext{4}{Universit\`a degli Studi di Trento, Dipartimento di Fisica, via Sommarive 14, I-38123 Trento, Italy}

\shorttitle{}

\email{* c.barbieri@campus.unimib.it}

\shortauthors{C.~Barbieri  et. al}


\begin{abstract}
In this letter we focus on the peculiar case of a coalescing compact-object binary whose chirp mass is compatible both with a neutron star-neutron star and black hole-neutron star system, with the black hole in the $\sim 3-5\msun$ range defined as the “mass gap". Some models of core-collapse supernovae predict the formation of such low-mass black holes and a recent observation seems to confirm their existence. Here we show that the nature of the companion to the neutron star can be inferred from the properties of the {\it kilonova} emission once we know the chirp mass, which is the best constrained parameter inferred from the gravitational signal in low-latency searches. In particular, we find that the kilonova in the black hole-neutron star case is far more luminous than in the neutron star-neutron star case, even when the black hole is non spinning. The difference in the kilonovae brightness arises primarily from the mass ejected during the merger. Indeed, in the considered interval of chirp masses, the mass ejection in double neutron star mergers is at its worst as the system promptly forms a black hole. Instead mass ejection for black hole-neutron star case is at its best as the neutron stars have low mass/large deformability. The kilonovae from black hole-neutron star systems can differ by two to three magnitudes. The outcome is only marginally dependent on the equation of state. The difference is above the systematics in the modeling.
\end{abstract}


\maketitle

\section{Introduction}
During the O1 and O2 observing runs, the LIGO Scientific Collaboration and Virgo Collaboration (LVC) detected gravitational wave (GW) signals from ten coalescing stellar-mass black hole binaries (BHBH) and a neutron star binary system (NSNS), the latter accompanied by a multi-wavelength electromagnetic (EM) counterpart \citep{GW170817,gw170817em,catalogo_GW}.
At the time of writing, as the third observing run (O3) is in progres,
the LVC reported the detection of two probable black hole-neutron star (BHNS) binary merger candidates (S190814bv -- \citealt{GCNS190814bv}, and S190910d -- \citealt{GCNS190910d}), plus candidates with a lower probability of being actual astrophysical events\footnote{We defer to the LIGO/Virgo O3 Public Alerts webpage \url{https://gracedb.ligo.org/superevents/public/O3/} for a complete list of current candidates.}.
Before the beginning of O3, the estimated BHNS detection rate for this run was in the range $0.04-12$ yr$^{-1}$ \citep{Dominik2015}. At the time of writing, there are no indications of observed EM counterparts associated with these candidates (\citealt{Coughlin2019_3}; for S190814bv see e.g. \citealt{Srivastav2019,Soares2019,Klotz2019}, for S190910d see e.g. \citealt{Crisp2019,Pereyra2019}).

On a theoretical point of view, BHNS mergers can be accompanied by an EM counterpart as in the NSNS case. This occurs when the NS is (at least partially) tidally disrupted before crossing the BH event horizon \citep{Shibata11}. Tidal disruption is favoured in binaries with low mass ratio $q=M_1/M_2$ and large NS tidal deformability $\Lambda_\mathrm{NS}$ (corresponding to a small NS mass and/or to a  ``stiff'' equation of state).  A high black hole spin\footnote{We use the term ``spin'' to indicate the dimensionless spin parameter, $\chi_\mathrm{BH}=cJ/GM_\mathrm{BH}^2$, where $J$ is the BH angular momentum.}, which brings the last stable circular orbit of the binary closer to the BH horizon, also greatly enhances the tidal disruption \citep{Shibata11,Kawaguchi2016,Foucart2018,Barbieri2019,Barbieri2019_2}. The unbound NS material (``ejecta'') is thought to produce kilonova emission \citep{Lattimer1974,Li1998,Metzger2017}. Moreover, \cite{Shapiro2017,Pasch2017,Ruiz-Jet-2018} showed that after a BHNS merger a relativistic jet can be launched, powering a short gamma-ray burst (sGRB) \citep{Eichler1989,Narayan1992} and GRB afterglow emission \citep{Sari1998,D'Avanzo2018,Ghirlanda2018,Salafia2019}.

The BH mass distribution observed so far in coalescing binaries is broad  \citep{Abbott18-10-bh}, extending up to $50^{+16.6}_{-10.2}\msun$, with the lightest BH carrying a mass $7.6^{+1.3}_{-2.1}\msun$, close to the mean BH mass in observed Galactic X-ray binaries of  $\sim7.8\pm1.2\msun$ \citep{Ozel2010}. Double NS systems observed so far carry masses in the interval $1.165\msun-1.590\msun$ \citep{Zhang2019}, and the NS with the highest and best estimated mass is the radio pulsar J0740+6620 with  $M_\mathrm{NS}=2.14^{+0.10}_{-0.09}\msun$ in a low-mass binary \citep{Cromartie2019}. Thus, observations appear to indicate  a discontinuity  between the observed mass distributions of NSs and stellar BHs, called “mass gap", located approximately 
between $\approx 3\msun$ (the maximum NS mass inferred from causality arguments)
and $\sim 5\msun$ \citep{Lattimer2001}. However recently \cite{Thompson2019} reported the discovery of a BH with mass $3.3^{+2.8}_{-0.7}\msun$ in a non-interacting binary system with a red giant.
 
The mass spectrum of compact objects depends sensitively on the 
mass of the carbon-oxygen core at the end of stellar evolution, on the compactness of the  collapsing core at bounce and on the supernova (SN) explosion engine. 
\cite{Belczynski2012} and \cite{Fryer2012} showed that, in the presence of a significant amount of fallback, explosions happening over a large interval of post-bounce times lead to a continuous range in remnant masses. By contrast explosions happening predominantly within a few hundreds of ms after bounce, characterized by negligible amounts of fallback material, produce more easily the mass gap. Interestingly, at the time of writing, the LVC reported event candidates with binaries having at least one component in the mass gap \citep{GCNS190924h,GCNS190930s}.

\begin{figure}
    \centering
    \includegraphics[width=\columnwidth]{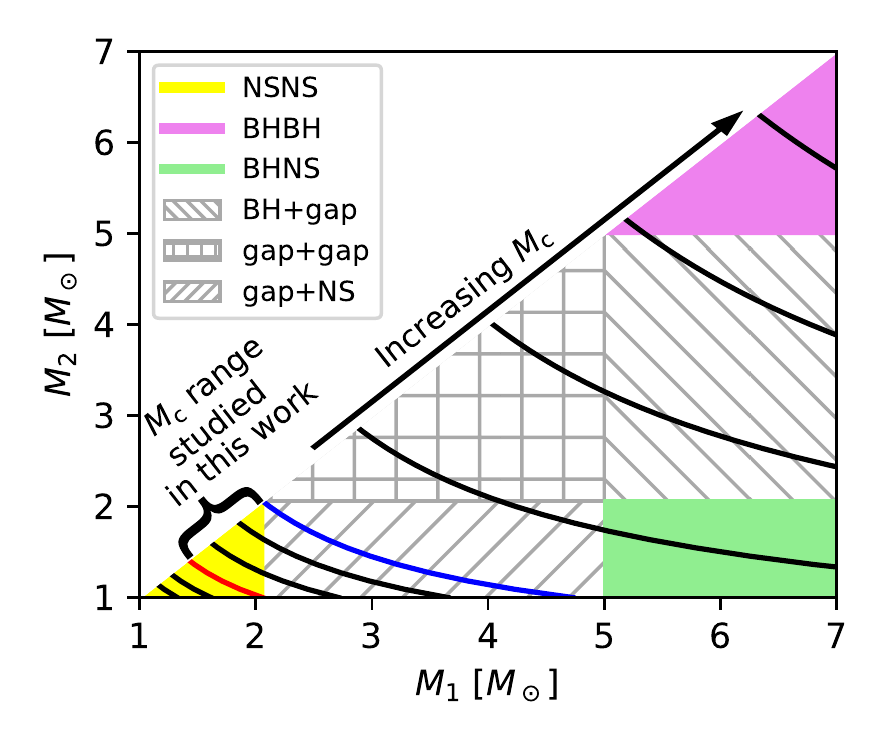}
    \caption{$M_1-M_2$ compact binaries having the same chirp mass $M_\mathrm{c}$. Different lines indicate different values for $M_\mathrm{c}$. We assume the SFHo EoS: the maximum NS mass is $M_\mathrm{NS}^\mathrm{max}=2.058\msun$, and  $M_\mathrm{NS}^\mathrm{min}$ is set equal to $1\msun$. Yellow, green and violet regions of the parameter space indicate, respectively, NSNS, BHNS and BHBH binaries. Gray-hatched areas indicate binary configurations with at least one component in the mass gap. Red and blue lines represent $M_\mathrm{c,min}^{\rm gapNS}$ and $M_\mathrm{c,max}^{\rm NSNS}$, respectively (see text for definition).}
    \label{fig:chirp_gap_nogap}
\end{figure}

It is known that the binary chirp mass $M_\mathrm{c}$, a combination of the masses of the two components, is one of the best measured parameters encoded in the GW signal. It is the prime parameter used to identify in low-latency searches the nature of the binary -- whether the system hosts two NSs, two stellar BHs or a BH and a NS. Interestingly, we note that if the NS and BH mass spectra join to form a continuum, i.e.~there is no ``mass gap'' between BH and NS mass distributions (as \citealt{Thompson2019} seem to indicate), there exists a range of values of the chirp mass $M_\mathrm{c}$ where the nature of the binary cannot be identified uniquely based on the chirp mass only \citep[see also][]{Mandel2015}.
In particular, hereafter we call ``ambiguous'' the chirp masses whose values are compatible with either a NSNS or a light BHNS system (see Fig.~\ref{fig:chirp_gap_nogap}).
 
In this Letter, we aim at answering the following question: can EM observations of coalescing binaries in this ``ambiguous'' chirp mass interval help to disentangle their nature and narrow down the uncertainties on the existence of a mass gap? To this purpose, we study the properties of the {\it kilonova} emission of NSNS and BHNS systems which fall in this ``ambiguous'' chirp mass interval, using the semi-analytical model presented in \cite{Barbieri2019,Barbieri2019_2}.

\section{``Ambiguous'' chirp masses}
\label{sec:degenerate_chirp}

The binary chirp mass is defined as
\begin{equation}
    M_\mathrm{c}=\frac{(M_1M_2)^{3/5}}{(M_1+M_2)^{1/5}},
\end{equation}
where $M_1$ and $M_2$ are the masses of the two component stars (we take $M_1\geq M_2$). LVC public alerts follow a classification scheme to communicate probabilistic estimates of the nature of the merging system to the community. The scheme classifies as ``BNS'' any system with both masses $M_1$ and $M_2$ smaller than $3\msun$; as ``BBH'' any system  with both $M_1$ and $M_2$ larger than $5\msun$; as ``NSBH'' any system with $M_1>5\msun$ and $M_2<3\msun$, and as ``MassGap'' any system with at least one component carrying a mass between $3$ and $5\msun$. An additional ``Terrestrial'' category is defined to represent triggers that are not of astrophysical origin (i.e.~false alarms). 

In this work we follow a slightly different classification. We assume the SFHo equation of state (EoS), for which the maximum  mass of a non-rotating NS is  $M_\mathrm{NS}^\mathrm{max}=2.058\msun$ \citep{SFHo}. We also fix the minimum NS mass to $M_\mathrm{NS}^\mathrm{min}=1\msun$ \citep[$\sim10\%$ lower than the value found in][]{Suwa2018}. We thus classify as ``NSNS'' those systems with both $M_1$ and $M_2$ between $1$ and $2.058\msun$ (yellow region in Fig.~\ref{fig:chirp_gap_nogap}); ``BHNS'' those with $M_1>5\msun$ and $M_2<2.058\msun$ (green region); ``BHBH'' those with both masses above $5\msun$ as in the LVC classification (purple region). Considering that compact objects populating the mass gap have masses larger than $M_\mathrm{NS}^\mathrm{max}$, we assume these to be stellar-origin BHs. In Fig.~\ref{fig:chirp_gap_nogap} we divide the ``MassGap'' region in three sub-regions: ``BH+gap'' for those systems with a BH above $5\msun$ and a BH in the gap; ``gap+gap'' for those with two BHs in the gap; ``gap+NS'' for those with a BH in the gap and a NS.

Two limiting values of the chirp mass can be identified: $M_\mathrm{c,min}^\mathrm{gapNS}=1.233\msun$ is the chirp mass corresponding to a gap+NS binary with $M_\mathrm{NS}=M_\mathrm{NS}^\mathrm{min}$ and $M_\mathrm{BH}=M_\mathrm{NS}^\mathrm{max}$ (red line). All GW events with chirp mass below $M_\mathrm{c,min}^{\rm gapNS}$ are NSNS mergers. Similarly, $M_\mathrm{c,max}^{\rm NSNS}=1.792\msun$ is the chirp mass corresponding to a NSNS binary with both NSs having the maximum allowed mass (blue line). Events with chirp mass above $M_\mathrm{c,max}^{\rm NSNS}$ cannot be produced by a NSNS merger. Events with chirp mass between $M_\mathrm{c,min}^{\rm gapNS}$ and $M_\mathrm{c,max}^{\rm NSNS}$ can be either NSNS or gap+NS mergers (green-orange lines),  i.e.~they are ``ambiguous''.

\section{Computation of ejecta properties from BHNS and NSNS mergers}
\label{sec:ejecta_properties}
During the final phase of a NSNS merger, tidal forces lead to a partial disruption of the stars, producing an outflow of neutron-rich material. When the crusts of the two NS impact each other, compression, shock heating and potentially neutrino ablation cause an additional outflow \citep{Hotokezaka2013,Bauswein2013,Radice2016,Dietrich2017,Beloborodov2018}. The released NS material can be divided into two components: the dynamical ejecta, gravitationally unbound, that leave the merger region, and a bound component, which forms an accretion disc around the merger remnant. On longer timescales, other outflows originate from the disc: faster ejecta produced by magnetic pressure and neutrino-matter interaction during the initial neutrino-cooling-dominated accretion phase (we call these ``wind ejecta''), and slower but more massive ejecta produced by viscous processes in the disc, especially during the advection-dominated phase (\citealt{Dessart2009,Metzger2014,Perego2014,Just2015,Siegel2017} -- we call these ``secular ejecta'').
Substantial differences in the ejecta properties arise depending on the post-merger scenario \citep[see i.e.][]{Kawaguchi2019}. 

In order to calculate dynamical ejecta and disc mass from a NSNS merger we adopt the fitting formulae reported in \cite{Radice2018_2}, which are calibrated on a large suite of high-resolution GRHD simulations\footnote{We note that \cite{Kiuchi2019} showed that the predictions from these formulae might underestimate the produced disc mass in binaries with large mass ratios. However they consider the case with $M_\mathrm{NS,1}=1.55\msun$ and $M_\mathrm{NS,2}=1.2\msun$, thus low-mass/largely deformable NSs. Instead, as can be seen in Fig. \ref{fig:dyn_disk}, we consider systems with $M_\mathrm{NS,1}>1.65\msun$ and $M_\mathrm{NS,2}>1.35\msun$. Therefore the NSs in our systems are less deformable and we expect that the underestimation reported in \cite{Kiuchi2019} is less significant.}. Both quantities depend on the NS masses and tidal deformabilities. We also adopt their formula for the dynamical ejecta mass-weighted average asymptotic velocity $v_\mathrm{dyn}$.

The NS tidal disruption can occur also in BHNS mergers.
If the NS is disrupted outside the innermost stable circular orbit, then the released material remains outside the BH in the form of a crescent \citep[e.g.][]{Kawaguchi2016}, otherwise the NS plunges directly onto the BH. We adopt the fitting formula from \cite{Foucart2018} to calculate the total mass remaining outside the BH, $M_\mathrm{out}$. This quantity depends on the BH mass and spin, and on the NS mass, tidal deformability and baryonic mass $M_\mathrm{NS}^\mathrm{b}$. We adopt the formulae in \cite{Kawaguchi2016} to calculate the dynamical ejecta mass and average velocity $v_\mathrm{dyn}$ in this case. $M_\mathrm{dyn}$ depends on the BH mass and spin, the NS mass, baryonic mass and compactness $C_\mathrm{NS}$, and on the angle $\iota_\mathrm{tilt}$ between the binary total angular momentum and the BH spin. We assume $\iota_\mathrm{tilt}=0$. $v_\mathrm{dyn}$ depends only on the mass ratio $q=M_\mathrm{BH}/M_\mathrm{NS}$. We proceed as in \cite{Barbieri2019} to calculate $C_\mathrm{NS}$ from $\Lambda_\mathrm{NS}$ and $M_\mathrm{NS}^\mathrm{b}$ from $M_\mathrm{NS}$ and $C_\mathrm{NS}$. We then calculate the disc mass as $M_\mathrm{disc}=\mathrm{max}[M_\mathrm{out}-M_\mathrm{dyn},0]$. As in \cite{Barbieri2019} we assume that $M_\mathrm{dyn}$ cannot exceed $30\%M_\mathrm{out}$, considering recent BHNS simulations presented in \cite{Foucart2019}.

In what follows, we conservatively assume the BH to be non-spinning ($\chi_\mathrm{BH}=0$), corresponding to the worst condition for ejecta production\footnote{among the co-rotating configurations. Indeed the counter-rotating cases ($\chi_\mathrm{BH}<0$) are the worst conditions in absolute, more often leading to a direct plunge. However counter-rotating configurations are not expected for field binaries but for the dynamically formed ones, that represent a negligible contribution to the merger rate \citep{Ye2019}.}.

\section{Ejecta properties for ``ambiguous'' chirp masses}
\label{sec:ejecta_properties_degenerate_chirp}

\begin{figure}
    \centering
    \includegraphics[width=\columnwidth]{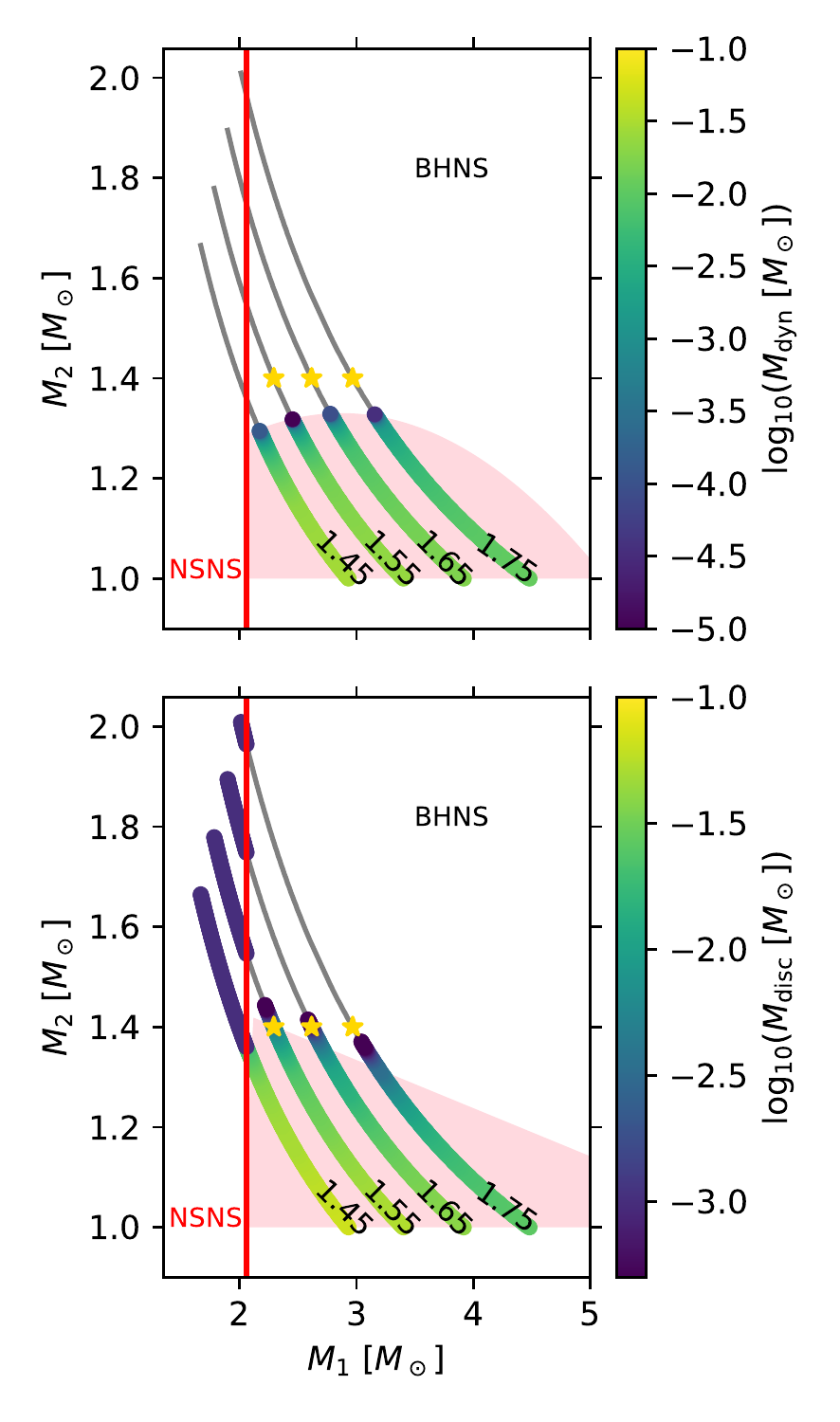}
    \caption{Dynamical ejecta (top panel) and accretion disc (bottom panel) masses for different values of the binary chirp mass $M_\mathrm{c}$. We assume the SFHo EoS ($M_\mathrm{NS}^\mathrm{max}=2.058\msun$) and non-spinning BHs ($\chi_\mathrm{BH}=0$). The vertical red line separates NSNS configurations (left) and BHNS ones (right). Each line corresponds to a $M_\mathrm{c}$ (reported on it). The pink shadowed area is the region where differences of the ejecta mass for the BHNS and NSNS cases are larger than systematic errors. The yellow stars indicate the BHNS systems with the NS having a “representative" mass of $1.4\msun$.}
    \label{fig:dyn_disk}
\end{figure}

\begin{figure*}
    \centering
    \includegraphics[width=\textwidth]{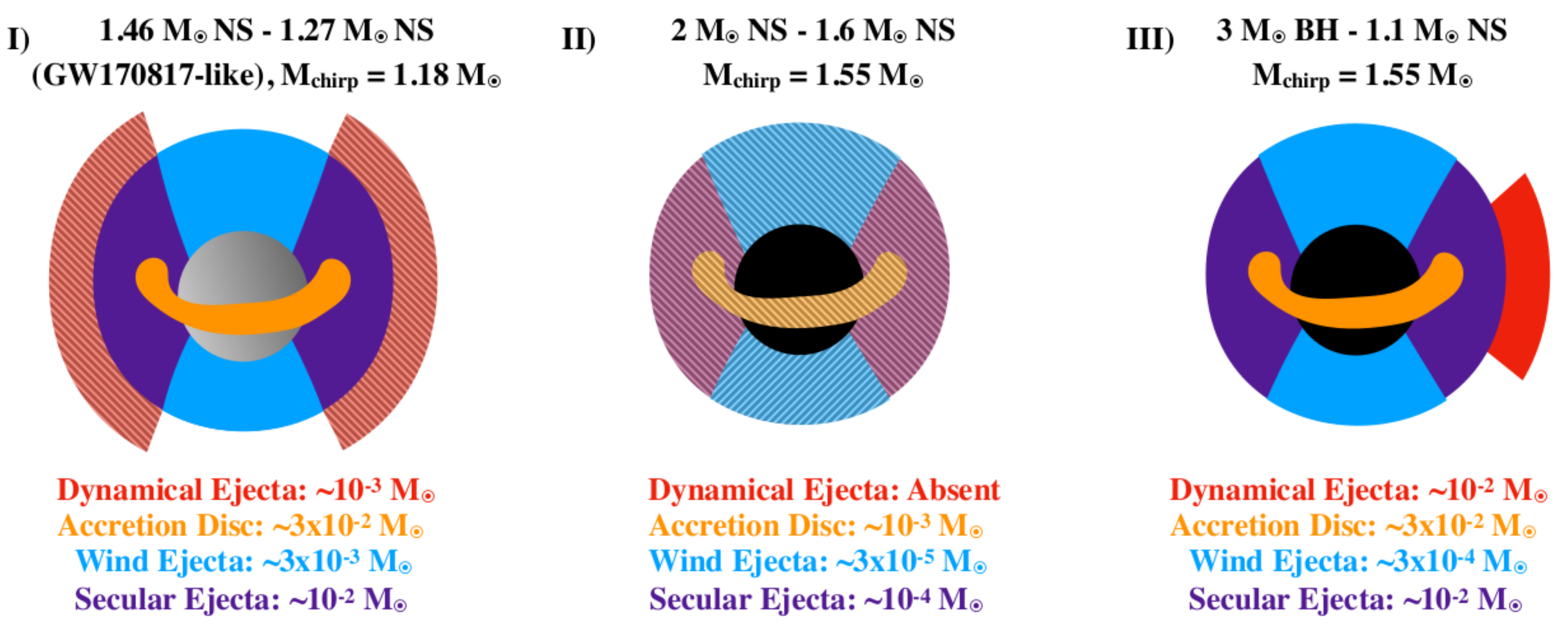}
    \caption{Cartoon of the ejecta and disc produced in different systems:  (I) a NSNS merger with $\sim 1.46\msun$ and $\sim 1.27\msun$ stars, close to the masses in GW170817; (II) a NSNS merger with two massive stars of $\sim2\msun$ and $\sim1.6\msun$; and (III) a BHNS merger with a light BH of $\sim3\msun$ and a NS of $\sim1.1\msun$. NSNS and BHNS configurations II and III correspond to the same ``ambiguous'' chirp mass $M_\mathrm{c}=1.55\msun$. Red, orange, light blue and purple represent dynamical ejecta, accretion disc, wind ejecta and secular ejecta, respectively. Filled areas correspond to massive components, while hatched areas correspond to low mass components.}
    \label{fig:confronto_ejecta}
\end{figure*}

In Fig.~\ref{fig:dyn_disk} we show the dynamical ejecta and disc masses on the $(M_1,M_2)$ plane along lines of constant $M_\mathrm{c}$. We limit the y axis to $M_\mathrm{NS}^\mathrm{max}$ as we focus on systems that contain at least one NS. 
It is apparent that NSNS configurations compatible with ``ambiguous'' chirp masses do not produce dynamical ejecta (upper panel of Fig.~\ref{fig:dyn_disk}). In this parameter region, the fits from \cite{Radice2018_2} predict the absence of this kind of ejecta, due to the prompt collapse of the remnant to a BH. Conversely, BHNS configurations can more easily produce dynamical ejecta. These systems have small mass ratio $q<5$ and low-mass (large $\Lambda_\mathrm{NS}$) NSs, which is the optimal condition to produce massive ejecta in BHNS mergers \citep[as shown in][]{Barbieri2019,Barbieri2019_2}. The same arguments hold for disc masses (bottom panel in Fig.~\ref{fig:dyn_disk}). Note that the value of $M_\mathrm{disc}$ predicted by the fitting formula for NSNS systems in the considered range is set by the lower limit indicated in \cite{Radice2018_2}, which is $M_\mathrm{disc}=10^{-3}$. For BHNS configurations, instead, discs with masses up to $\sim7\times10^{-2}\msun$ are produced.

It is important to note that when the differences of the ejecta mass for the BHNS and NSNS cases are substantial, they are larger than the systematic errors. The uncertainties on the fitting formulae for NSNS are $\Delta M_\mathrm{dyn}^\mathrm{NSNS}=0.5~M_\mathrm{dyn}+5\times10^{-5}\msun$ and $\Delta M_\mathrm{disc}^\mathrm{NSNS}=0.5~M_\mathrm{disc}+5\times10^{-4}\msun$ \citep{Radice2018_2}. The uncertainties on the fitting formulae for BHNS are $\Delta M_\mathrm{dyn}^\mathrm{BHNS}=0.2~M_\mathrm{dyn}$ \citep{Kawaguchi2016} and $\Delta M_\mathrm{out}=0.1~M_\mathrm{out}$ \citep{Foucart2018}. Therefore, being $M_\mathrm{disc}^\mathrm{BHNS}=M_\mathrm{out}-M_\mathrm{dyn}^\mathrm{BHNS}$, we assume its uncertainty to be $\Delta M_\mathrm{disc}^\mathrm{BHNS}=\sqrt{(\Delta M_\mathrm{out})^2+(\Delta M_\mathrm{dyn}^\mathrm{BHNS})^2}$. We define $\sigma_\mathrm{dyn}=\sqrt{(\Delta M_\mathrm{dyn}^{\rm NSNS})^2+(\Delta M_\mathrm{dyn}^{\rm BHNS})^2}$ and $\sigma_\mathrm{disc}=\sqrt{(\Delta M_\mathrm{disc}^{\rm NSNS})^2+(\Delta M_\mathrm{disc}^{\rm BHNS})^2}$. We indicate as pink shadowed area in Fig. \ref{fig:dyn_disk} the regions where the differences in the mass of dynamical ejecta and disc for the BHNS and NSNS cases are greater than or equal to $\sigma_\mathrm{dyn}$ and $\sigma_\mathrm{disc}$, respectively. In these regions the ejecta mass differences are larger than the systematic errors.

Figure~\ref{fig:confronto_ejecta} summarizes the differences between two representative NSNS and BHNS systems with ``ambiguous'' chirp masses (cases II and III in the Figure), and also a ``GW170817-like'' NSNS case, for comparison. For the latter we consider a NSNS system with masses $1.46\msun$ and $1.27\msun$. 

Merger (I) produces relatively low-mass dynamical ejecta at all latitudes, with a preferentially equatorial angular distribution $\propto \mathrm{sin}^2\theta$, where $\theta$ is the polar angle \citep{Perego2017}. The accretion disc is massive and $\sim20\%$ of its mass is unbound in the form of secular ejecta, with a similar angular distribution as for the dynamical ones, while $\sim5\%$ of its mass goes into the wind ejecta, mostly confined in the polar region ($\theta<\pi/3~\mathrm{rad}$ -- \citealt{Perego2017}). After the merger, an intermediate state with a hyper-massive NS could exist before collapsing to a BH (gray central object represented in Fig.~\ref{fig:confronto_ejecta}-I). The strong neutrino winds produced in this state interact with the ejecta, increasing the electron fraction $Y_\mathrm{e}$ or, equivalently, lowering the opacity. 

We consider a system with $2\msun$ and $1.6\msun$ stars (II) as our representative NSNS merger in the ``ambiguous'' chirp mass range. As explained above, in this case we expect no dynamical ejecta and a low-mass accretion disc, resulting in low-mass wind and secular ejecta. The merger remnant collapses promptly to a BH. The absence of an intermediate hyper-massive NS state implies little neutrino wind, giving a low $Y_\mathrm{e}$ in the ejecta \citep{Kawaguchi2019}.

Finally, as BHNS merger in the ``ambiguous'' chirp mass range we consider a system with $M_\mathrm{BH}=3\msun$ and $M_\mathrm{NS}=1.1\msun$ (III). In BHNS mergers, the dynamical ejecta have a crescent-like shape, extending into half of the equatorial plane and limited to the region with $\theta<0.3~\mathrm{rad}$ \citep{Kawaguchi2016}. In the considered system, the dynamical ejecta and accretion disc are massive. Due to the lack of a neutrino wind, the fraction of accretion disc flowing into wind ejecta is lower than in the NSNS case (we assume $\sim1\%$). The disc fraction that goes into secular ejecta is the same as in the NSNS case. As a consequence, the secular ejecta are massive, while the wind ejecta have low mass. The ejecta $Y_\mathrm{e}$ is lower than the NSNS case.

Therefore, being the ejecta properties substantially different for the NSNS and BHNS cases in the ``ambiguous'' chirp mass range, we expect the kilonova light curves to present important differences as well.

\section{Kilonova model}
\label{sec:kilonova_model}
The neutron-rich material ejected in NSNS and BHNS mergers is an ideal site for $r$-process nucleosynthesis, which produces the heaviest elements in the Universe \citep{Lattimer1974,Eichler1989,Korobkin2012,Wanajo2014}. The synthesized nuclei are unstable and they decay radioactively, powering the kilonova emission \citep{Li1998,Metzger2010,Kasen2013}.

We compute the kilonova light curves using the composite semi-analytical model presented in \cite{Barbieri2019,Barbieri2019_2} \citep[in part based on][]{Perego2017,Martin2015,Grossman2014}. For the NSNS cases we assume the model parameters (ejecta geometry, opacity and the fractions of $M_\mathrm{disc}$ that go into wind and secular ejecta) as in \cite{Perego2017}. The model has been tested on GW170817: using the parameters inferred for this event \citep{Perego2017,GW170817_dynamical}, we obtain multi-wavelength light curves consistent with the observed ones \citep[paper in preparation]{Villar2017}. For BHNS systems we assume the same model parameters as in \cite{Barbieri2019,Barbieri2019_2} \citep[based on][]{Kawaguchi2016,Fernandez2017,Just2015}. 

The kilonova light curves are highly degenerate with respect to binary parameters. Thus, it is impossible to infer the system properties from the kilonova light curve alone. This degeneracy can be (at least partially) broken using information from GW analysis. In particular, the measurement of the binary chirp mass reduces the number of parameters by one. Leaving i.e.~$M_1$ as a free parameter, $M_2$ is constrained by the measured $M_\mathrm{c}$.

\section{Kilonovae for ``ambiguous'' chirp masses}
\label{sec:kilonova_degenerate_chirp}

\begin{figure*}
    \centering
    \includegraphics{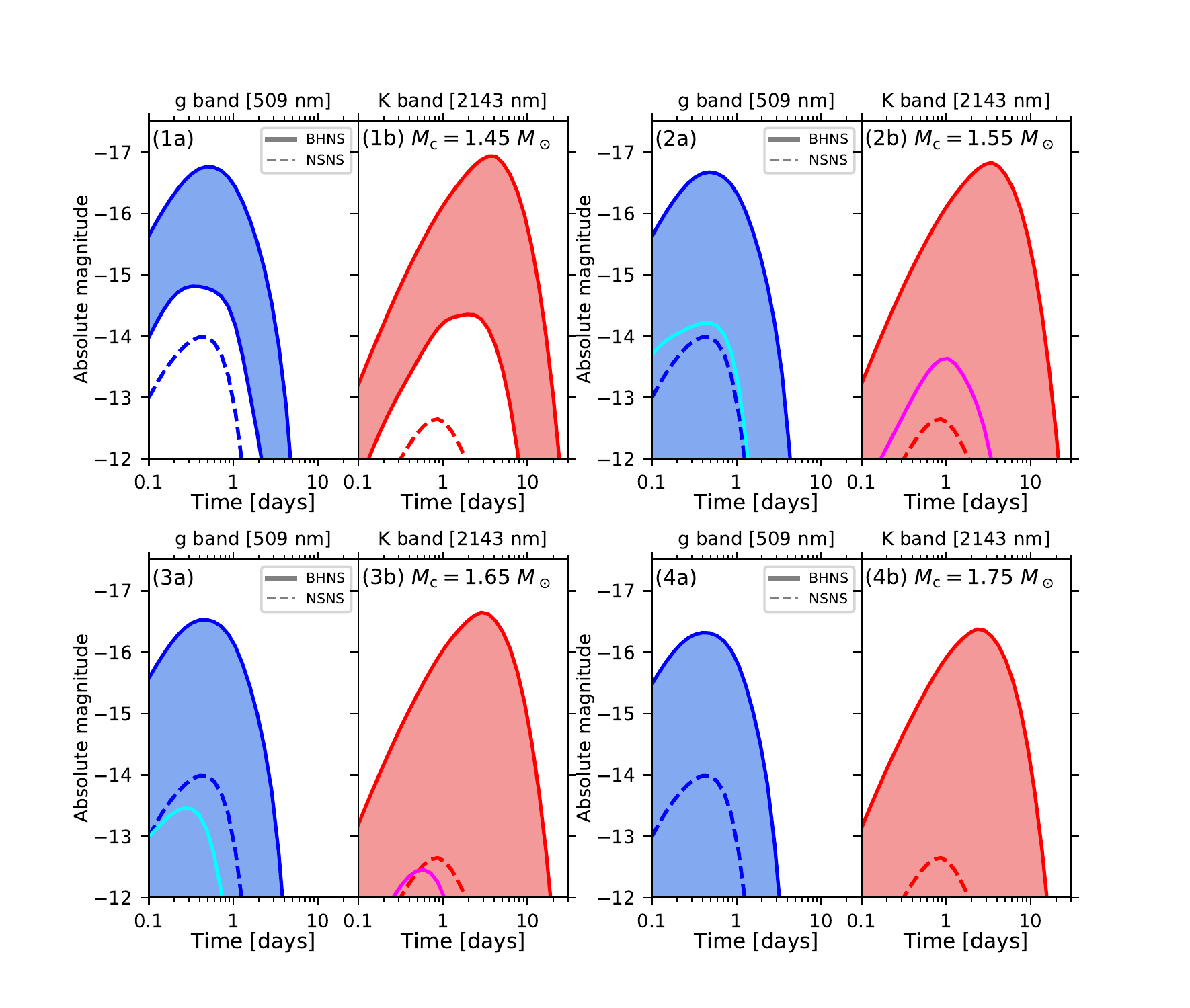}
    \caption{Light curve ranges for kilonovae produced by NSNS (dashed lines) and BHNS (filled areas) binaries with a chirp mass $M_\mathrm{c}=$ $1.45\msun$ (1), $1.55\msun$ (2), $1.65\msun$ (3) and $1.75\msun$ (4). The “a" and “b" panels show light curve ranges in the g (509 nm) and K (2143 nm) band, respectively. Aqua (magenta) lines represent the kilonova in the g (K) band for the BHNS systems with a NS having a “representative“ mass of $1.4\msun$.}
    \label{fig:kilonovae}
\end{figure*}

In Fig.~\ref{fig:kilonovae} we show the envelope of the kilonova light curves expected from NSNS and BHNS mergers, for four selected values of the chirp mass. We consider emission in the g (509 nm) and K (2143 nm) bands and the figure shows the absolute magnitude as a function of time. For all the ``ambiguous'' chirp masses the fitting formulae from \cite{Radice2018_2} in NSNS mergers predict no dynamical ejecta and a minimum allowed disc mass $M_\mathrm{disc}=10^{-3}\msun$. Thus we have a single light curve for NSNS mergers, and we can expect that these events would not produce kilonovae brighter than shown in  Fig.~\ref{fig:kilonovae}. 

For BHNS mergers there exists a range of light curves for each $M_\mathrm{c}$, arising from the different combinations of the component masses, producing different ejecta properties.
For $M_\mathrm{c}=1.45\msun$ (panels 1a-1b) all kilonovae from BHNS mergers are much brighter at every time than that from NSNS mergers. Therefore a single observation in one of these bands  would allow in principle  to distinguish the nature of the merging system.

At higher values of the chirp mass, there is only a small overlap between the BHNS and NSNS cases, at the bottom of the BHNS envelope. Therefore, except for observed light curves at low absolute magnitudes, it should be always possible to distinguish the nature of the merging system by the sole kilonova brightness. We note that the disentangling of the nature of the binary is optimal when $M_\mathrm{c}=1.45\msun$ (panels 1a-1b). In this case, as shown in Fig.~\ref{fig:dyn_disk}, the mass interval of the ejecta from BHNS mergers is the narrowest, and this in turn leads to the narrowest spread in the kilonova light curves.

The prediction of BHNS kilonovae as bright or brighter than NSNS ones is presented in \cite{Kawaguchi2019}. They find that BHNS kilonovae are brighter in the near-infrared K-band, due to the smaller electron fraction $Y_\mathrm{e}$ in the ejecta owing to the lack of strong neutrino irradiation from the central remnant. In the i band, \cite{Kawaguchi2019} find that NSNS configurations ending with the formation of a supermassive NS leads to brighter kilonovae than the BHNS case. This is due to the strong neutrino emission produced in this case, that increases $Y_\mathrm{e}$ in the ejecta. However, in their study they compare sundry BHNS and NSNS configurations not selected on the bases of the chirp mass. By contrast, in our work, we compare BHNS and NSNS mergers at fixed chirp mass. This requirement restricts the NSNS binary configurations to cases producing no dynamical ejecta and very low mass discs. Therefore, whatever the value of $Y_\mathrm{e}$ in the ejecta from NSNS merger is, the mass propelled in the merger is so low that almost all the BHNS kilonovae are brighter, at all wavelengths. 

For other comparisons between NSNS and BHNS merger outcomes and studies on distinguishing the nature of merging compact binaries see \cite{Hinderer2019,Coughlin2019}, who considered an unconventional BH companion with mass of $\sim1.4\msun$, thus below the maximum NS mass. 

As a visual comparison we also show the kilonovae for BHNS binaries having a NS with a “representative" mass of $1.4\msun$ (aqua/magenta lines). For $M_\mathrm{c}=1.45\msun$ such a binary does not exist, while for $M_\mathrm{c}=1.75\msun$ it is fated to a direct plunge, thus there is no kilonova.

We remark that the kilonova light curves from BHNS are inferred assuming non-spinning BHs ($\chi_\mathrm{BH}=0$). 
As explained in \cite{Barbieri2019,Barbieri2019_2}, increasing the BH spin (fixing all the other parameters) leads to more massive ejecta and, consequently, more luminous kilonovae. Therefore, if the BHs have a non-zero spin, our argument would be even stronger. As an example, for $\chi_\mathrm{BH}\gtrsim0.5$ all light curves from BHNS kilonovae would be brighter than those from NSNS binaries in each band and at any time, in this critical range of “ambiguous" chirp masses.

\section{Conclusion}

The detection of a BHNS coalescence could be the next 
ground-breaking discovery in multi-messenger astronomy. At the time of 
writing, there are promising GW candidates detected by the LVC during the observation run O3. The associated detection of an electromagnetic signal from these
new GW sources might contribute to our understanding of the physical processes that power  
the multi-wavelength EM emission \citep[][and references therein]{Gompertz2018,Rossi2019}.

From the GW signal, one of the best constrained parameters in low latency is the binary chirp mass, a combination of the masses of the two components. This parameter is currently used to classify the binary, whether the system hosts two NSs, two stellar BHs or a BH and a NS. 
In the present Letter, we point out that in absence of a ``mass gap'' between the NS and BH mass distributions (as \citealt{Thompson2019} seem to indicate), there exists a range of $M_\mathrm{c}$ (as shown in Fig.~\ref{fig:chirp_gap_nogap}) for which it is not possible to distinguish the nature of the binary on the basis of the chirp mass measurement alone\footnote{Offline GW signal analysis could provide more precise information. This would lead to stronger constraints on the component masses that could allow to distinguish the nature of the merging system using GW alone, in some cases.}.
For the SFHo EoS adopted in this analysis, we find that the values of the chirp mass between $1.233\msun$ and $1.792\msun$ are compatible either with NSNS and BHNS systems. 
 
In this Letter we show that the observation of the kilonova emission from these systems can break the degeneracy in the ``ambiguous'' chirp mass range, and constrain the nature of the merging system.
We find that kilonova emission shows substantial differences in the luminosity and temporal evolution in NSNS and BHNS systems (see Fig.~\ref{fig:kilonovae}). In particular, the BHNS case is far more luminous than the NSNS case, even when the BH is non-spinning. This happens because in in this ``ambiguous'' $M_\mathrm{c}$ range the NSNS configurations represent the worst cases for ejecta production, while the BHNS configurations represent the best ones. It is important to note that, when the differences of the ejecta mass for the BHNS and NSNS cases are substantial, they are larger than the systematic errors in the modeling. Therefore, observing the kilonova associated with such an event is of fundamental importance to break the degeneracy on the nature of the merging system. Furthermore, if the observed kilonova is compatible with a BHNS merger, this would provide evidence in support of the existence of low-mass BHs, filling the ``mass gap''. 

This work illustrates the potential of multi-messenger observations of compact binary mergers, and the importance of an efficient exchange of information between the GW and EM communities \citep{Biscoveanu2019,Margalit2019ApJ}.

\begin{acknowledgements}
\scriptsize{We thank F.~Zappa and S.~Bernuzzi for sharing EoS tables. The authors acknowledge support from INFN, under the Virgo-Prometeo initiative. O.~S. and G.G. acknowledges the Italian Ministry for University and Research (MIUR) for funding through project grant 1.05.06.13.}
\end{acknowledgements}

\bibliography{references.bib}
\end{document}